\makeatletter

\def\@sect#1#2#3#4#5#6[#7]#8{\ifnum #2>\c@secnumdepth
     \def\@svsec{}\else
     \refstepcounter{#1}\edef\@svsec{\csname the#1\endcsname.\hskip 1em }\fi
     \@tempskipa #5\relax
      \ifdim \@tempskipa>\z@
        \begingroup #6\relax
          \@hangfrom{\hskip #3\relax\@svsec}{\interlinepenalty \@M #8\par}
        \endgroup
       \csname #1mark\endcsname{#7}\addcontentsline
         {toc}{#1}{\ifnum #2>\c@secnumdepth \else
                      \protect\numberline{\csname the#1\endcsname}\fi
                    #7}\else
        \def\@svsechd{#6\hskip #3\@svsec #8\csname #1mark\endcsname
                      {#7}\addcontentsline
                           {toc}{#1}{\ifnum #2>\c@secnumdepth \else
                             \protect\numberline{\csname the#1\endcsname}\fi
                       #7}}\fi
     \@xsect{#5}}

\def\label#1{\@bsphack\if@filesw {\let\thepage\relax
   \xdef\@gtempa{\write\@auxout{\string
      \newlabel{#1}{{\thesection.\@currentlabel}{\thepage}}}}}\@gtempa
   \if@nobreak \ifvmode\nobreak\fi\fi\fi\@esphack}

\def\@eqnnum{(\thesection.\theequation)}
\documentstyle[12pt]{article}
\makeatletter
\def\section{\setcounter{equation}{0} \@startsection {section}{1}{\z@}{-3.5ex
   plus -1ex minus -.2ex}{2.3ex plus .2ex}{\Large\bf}}
\def\l{\lambda}

\def\g{\gamma}
\def\mw{m_W}

\def\Im{{\cal I \mskip-4.5mu \lower.1ex \hbox{\it m}}\,}
\def\Re{{\cal R \mskip-4mu \lower.1ex \hbox{\it e}}\,}
\def\IJMP #1 #2 #3 {{\it Int.\ J.\ Mod.\ Phys.}\ {\bf #1}\ (#2) #3}
\def\MPL #1 #2 #3 {{\it Mod.\ Phys.\ Lett.}\ {\bf #1}\ (#2) #3}
\def\NC #1 #2 #3 {{\it Nuovo Cim.}\ {\bf #1} (#2) #3}
\def\NP #1 #2 #3 {{\it Nucl.\ Phys.}\ {\bf #1}\ (#2) #3}
\def\PL #1 #2 #3 {{\it Phys.\ Lett.}\ {\bf #1}\ (#2) #3}
\def\PR #1 #2 #3 {{\it Phys.\ Rev.}\ {\bf #1}\ (#2) #3}
\def\PP #1 #2 #3 {{\it Phys.\ Rep.}\ {\bf #1}\ (#2) #3}
\def\PRL #1 #2 #3 {{\it Phys.\ Rev.\ Lett.}\ {\bf #1}\ (#2) #3}
\def\RMP #1 #2 #3 {{\it Rev.\ Mod.\ Phys.}\ {\bf #1}\ (#2) #3}

\author{G. Jikia$^{(a)}$  and  A. Tkabladze$^{(b)}$ \\
[1ex]{${}^{(a)}$\it Institute for High Energy Physics},\\
     {\it 142284, Protvino, Moscow Region,}\\
     {\it Russian Federation}\\
[1ex]{${}^{(b)}$\it Kutaisi State University}\\
{\it 384000, Kutaisi, Georgia} }
\title{{\large \hfill IHEP 93--151\\ \hfill hep-ph/9312274\\
	\hfill December 1993 \\ \vspace*{3cm}}
	$\g Z$ Pair Production at the Photon Linear Collider}
\date{}
\begin{document}
\maketitle

\begin{abstract}
$\g\g\rightarrow\g Z$ scattering at the Photon Linear Collider is considered.
Explicit formulas for helicity amplitudes due to $W$ boson loops are presented.
It is shown that the $Z\g$ pair production will be easily observable at PLC
and separation of the $W$ loop contribution will be possible at $e^+e^-$ c.m.
energy of 300~GeV or higher.
\end{abstract}

\newpage

\section{Introduction}
The coupling of the three photons to $Z$-boson is absent at the classical level
and is generated only at the one loop order. The study of this coupling gives
us a possibility to probe nonabelian structure of the Electroweak Interactions,
through  the $W$ boson loop contribution involving both triple and quartic
bosonic vertices, in addition to the fermionic loop diagrams.

Until recently a special attention was attracted by the decay $Z\to\g\g\g$,
motivated by $LEP$ experiments. The decay width was calculated taking into
account both fermion \cite{fermi} and $W$ \cite{dong,GM} loop contributions.
The conclusion is that the $Z\to\g\g\g$ decay width is too small  and is
clearly beyond even the high luminosity option of $LEP$.

With the advent of the new collider technique \cite{plc} a more favorable
opportunity for the experimental study of the $\g\g\g Z$ coupling appears. It
is the scattering reaction $\g\g\to\g Z$ in the collision of high energy high
intensity photon beams at the Photon Linear Collider (PLC).

The $W$ loop contribution to the polarization tensor of the process
$Z\to\g\g\g$ was derived recently in \cite{GM}, where the decay width was
calculated.  To minimize the number of the Feynman diagrams all calculations
were done in unitary gauge, were the individual diagrams contain superficial
divergencies which were canceled before reducing the tensor integrals to scalar
integrals.

We have derived the helicity amplitudes for $\g\g\to\g Z$ in 't~Hooft-Feynman
and non-linear gauges \cite{gauge}. In both these gauges superficial
divergencies  do not appear. In 't~Hooft-Feynman gauge there are much more
diagrams due to the triple $W$-Nambu-Goldstone-photon ($Z$ boson) couplings
than in unitary gauge.  However, in non-linear gauge, where such mixed coupling
do not appear, there are just diagrams with either $W$ boson, NG or ghost
fields inside the loop.

In Section~2 we present explicit analytic expressions for the $W$-boson loop
contribution to the helicity amplitudes. Section~3  contains the numerical
results. It is shown that $\g\g\to\g Z$ cross section is large enough to be
observable at the PLC and even separation of the $W$ loop should be possible at
large energy. This fact has a fundamental significance for testing triple and
quartic $W$ boson vertices.  Finally, in Section~4 conclusions are made.

\section{Helicity Amplitudes}

We use the reduction algorithm of \cite{olden} to express the
helicity amplitudes of the reaction $$\g(p_1)\g(p_2)\to\g(p_3) Z(p_4)$$ in
terms of the set of basic scalar loop integrals. As noted above, all
calculations were done in both 't~Hooft-Feynman and non-linear gauges.  The
algebraic calculations were carried out using symbolic manipulation program
FORM \cite{FORM}.

Leaving out the factor $\alpha^2 \cos\theta_W/\sin\theta_W$, we find the nine
independent amplitudes:

\begin{eqnarray}
\lefteqn{A_{++++}(s,t,u) =  \frac{16 s_1}{s}E(t,u)+} \nonumber \\
&&4 (2(s-4 m_W^2)s_1-m_W^2(m_Z^2-6 m_W^2))
[D(s,t)+D(s,u)+D(t,u)] + \nonumber \\
&&2\biggl(\frac{m_Z^2}{m_W^2}-6\biggr)\Biggl\{\frac{t u +
m_W^2(s_1+s)}{s s_1} E(t,u) -
\frac{2 m_W^2}{s_1}[t u D(t,u) +m_Z^2 C(s)]-\nonumber \\
&&\frac{(s+m_Z^2)t u}{s_1 t_1 u_1}-\biggl[\frac{2 m_W^2 m_Z^2 s}{s_1 t_1}
 C_1(t)-\biggl(\frac{2 t+s}{s_1} - \frac{m_Z^4 s}{s_1 t_1^2}\biggr) B_1(t) +
(u\leftrightarrow t)\biggr]\Biggr\}.
\label{apppp}
\end{eqnarray}

\begin{eqnarray}
\lefteqn{A_{+++-}(s,t,u) = }\nonumber \\
&&2\biggl(\frac{m_Z^2}{\mw^2}-6\biggr)
\Biggl\{-2m_W^4(D(s,t)+D(s,u)+D(t,u))
- \frac{m_Z^2 t u}{s^2 s_1} E(t,u) + \nonumber\\
&&m_W^2\biggl(\frac{4m_Z^2-s}{s s_1} D(t,u)-\frac{s(u^2+t^2)}{s_1 t u} C(s)-
\frac{s_1^2}{u t} C_1(s)\biggr) +
\frac{(s+m_Z^2) u t}{s_1 t_1 u_1}+ \nonumber\\
&&\biggl[m_W^2 \biggl(
\biggl(\frac{(m_Z^2 u-s t) s}{s_1 t_1 u}+\frac{2 m_Z^2 u -s u_1}{s_1 s}\biggr)
C_1(t)-\frac{(2m_Z^2 u+s t)t}{s u s1} C(t)- \nonumber         \\
&&\frac{s t}{u} D(s,t)\biggr)+\frac{m_Z^2 (2 t_1-s) u t}{s s_1 t_1^2} B_1(t)+
(u \leftrightarrow t)\biggr]
\Biggr\}.
\label{apppm}
\end{eqnarray}

\begin{eqnarray}
\lefteqn{ A_{++-+}(s,t,u) = 2\biggl(\frac{m_Z^2}{m_W^2}- 6\biggr)
\Biggl\{-2 m_W^4[D(s,t)+D(s,u)+D(t,u)] -} \nonumber\\
&&m_W^2\biggl(\frac{t u}{s_1} D(t,u) +\frac{s(u^2+t^2)}{s_1 t u}C(s)+
\frac{s_1^2}{u t}C_1(s)\biggr)+1-\nonumber\\
&&\biggl[m_W^2\biggl(\frac{s t}{u} D(s,t) +\frac{u^2+t^2}{s_1 u}C(t)+
\frac{t t_1}{s_1 u}C_1(t)\biggr) + (u \leftrightarrow t)\biggr]\Biggr\}.
\label{appmp}
\end{eqnarray}

\begin{eqnarray}
A_{++--}(s,t,u)& =& 2\biggl(\frac{m_Z^2}{m_W^2} - 6\biggr)
\biggl\{1-2m_W^4[D(s,t) + D(s,u) +D(t,u)] - \nonumber\\
&&\frac{m_W^2 m_Z^2}{s s_1}[E(u,t) + 2 s C(s)]\biggr\}.
\label{appmm}
\end{eqnarray}

\begin{eqnarray}
\lefteqn{ A_{+-+-}(s,t,u) = A_{+--+}(s,u,t) =
 \frac{16 s}{s_1} E(s,u)+}\nonumber\\
&&4 \biggl(\frac{2 s t(t-4 m_W^2)}{s_1}-m_W^2(m_Z^2-6 m_W^2)\biggr)
[D(s,t)+D(s,u)+D(t,u)] + \nonumber \\
&& 2\biggl(\frac{m_Z^2}{m_W^2}-6\biggr)\Biggl\{\biggl(\frac{s u}{t^2}+
\frac{2 m_W^2}{t}\biggr)
E(s,u)-m_W^2\biggl(\frac{2 s u}{t} D(s,u)+\frac{m_Z^2}{s s_1} E(t,u)+
\nonumber\\
&&\frac{2 m_Z^2(2u_1+s)}{s_1 u_1} C_1(u)\biggr)+ \frac{s(s_1-u)}{s_1 t} B_1(s)-
\frac{s u (2s_1 u_1 +t u)}{s_1 t u_1^2}B_1(u) - \frac{s u}{s_1 u_1}\Biggr\}.
\label{apmpm}
\end{eqnarray}

\begin{eqnarray}
\lefteqn{ A_{+---}(s,t,u) = A_{+-++}(s,u,t) = }\nonumber\\
&&\frac{8 m_Z^2 t}{s_1 u}(2 E(s,t)-
u(4 m_W^2 - u)(D(s,t) + D(s,u) +D(t,u)) - \nonumber \\
&&2\biggl(\frac{m_Z^2}{m_W^2}-6\biggr)\Biggl\{m_W^2\biggl[\biggl(\frac{s t}{u}
+ 2 m_W^2\biggr) D(s,t)
+\biggl(\frac{s u}{t} +2 m_W^2\biggr)D(s,u)+
 \nonumber \\
&&\biggl(\frac{t u}{s_1}+2 m_W^2\biggr)D(t,u) +
 \frac{s s_1}{u t} C(s) +\frac{t^2}{s_1 u}C(t) +\frac{t^2+s_1^2}{s_1 t} C(u)+
 \nonumber \\
&&\frac{u^2+t^2}{u t}C_1(s)+\frac{2 m_Z^2 u^2 +t_1(u u_1+s s_1)}
{s_1 t_1 u}C_1(t)+
\frac{u u_1}{s_1 t}C_1(u)\biggr]+\nonumber\\
&&\frac{m_Z^2 t u}{s_1 t_1^2} B_1(t) - \frac{s t}{s_1 t_1}\Biggr\}
\label{apmmm}
\end{eqnarray}

\begin{eqnarray}
\lefteqn{ A_{+++0}(s,t,u)/(p_t/\sqrt{2}) =
 2 m_Z\biggl(\frac{m_Z^2}{\mw^2}-6\biggr)}\nonumber\\
&&\Biggl\{(t-u)\biggl(\frac{3 m_W^2}{s_1} D(t,u)-
\frac{E(t,u)}{s s_1} + \frac{2 m_W^2 s}{s_1 t u} C(s) +\frac{2 s}{s_1 t_1 u_1}
\biggr)+ \nonumber \\
&&\biggl[m_W^2\biggl(\frac{s}{u} D(s,t) +\frac{2}{s_1}C(t)+
2\frac{2 s^2-t_1^2}{s_1 t_1 u}C_1(t)\biggr)+\frac{2 t_1 t +m_Z^2 u}{s_1 t_1^2}
B_1(t) - \nonumber\\
&&(u\leftrightarrow t)\biggr]\Biggr\}.
\label{appp0}
\end{eqnarray}

\begin{eqnarray}
 A_{++-0}(s,t,u)/(p_t/\sqrt{2}) = 2 m_Z(m_Z^2 - 6m_W^2)
\Biggl\{\frac{(t-u)}{s_1}\biggl(D(t,u) - \frac{2 s}{u t}\biggr)-
\nonumber \\
 \biggl[\frac{s}{u}D(s,t) - \frac{2}{s_1}C(t)+\frac{2 t_1}{s_1 u} C_1(t)-
(u\leftrightarrow t)\biggr]\Biggr\}.
\label{appm0}
\end{eqnarray}

\begin{eqnarray}
\lefteqn{ A_{+-+0}(s,t,u)/(p_t/\sqrt{2}) = A_{+--0}(s,u,t)/(p_t/\sqrt{2})=}
\nonumber \\
&&-\frac{16 m_Z s}{s_1}\biggl((t-4 m_W^2)(D(s,t)+
D(s,u) + D(t,u))+\frac{2 E(s,u)}{t}\biggr) - \nonumber \\
&& 2 m_Z\biggl(\frac{m_Z^2}{m_W^2}- 6\biggr)\Biggl\{m_W^2\biggl(
\frac{(u -t)}{s_1} D(t,u) +
\frac{s}{u}D(s,t) +\frac{3 s}{t} D(s,u)\biggr)+\nonumber\\
&&\frac{2 m_W^2}{s_1}\biggl(C(u)-C(t)-\frac{s s_1}{u t} C(s)+\frac{t_1}{u}
C_1(t) + \frac{u_1}{t}C_1(u)-\frac{2 t}{u_1} C_1(u)\biggr)\nonumber \\
&& -\frac{s}{t^2}E(s,u)+\frac{2 s}{s_1 t} B_1(s)- 2\biggl(\frac{1}{t} +
 \frac{m_Z^2 t}{s_1 u_1^2}\biggr)
B_1(u) + \frac{2 s}{s_1 u_1}\Biggr\}.
\label{apmp0}
\end{eqnarray}

The other helicity amplitudes can be obtained by parity transformation.
We define
$$
s = (p_1+p_2)^2,~~~~ t = (p_2+p_3)^3,~~~~ u = (p_1+p_3)^2.
$$
All amplitudes with longitudinal polarization of $Z$-boson contain
$p_t$ -- transverse momentum of final particles
$$
p_t = \sqrt{\frac{t u}{s}}.
$$
The scalar three-point functions are given by
\begin{eqnarray}
C(s) =
\frac{1}{i\pi^2}\int\frac{d^4q}{\left(q^2-m_W^2\right)
\left((q+p_1)^2-m_W^2\right)\left((q+p_1+p_2)^2-m_W^2\right)}, \dots
\end{eqnarray}
and other two-, three- and four-point functions $B$, $C$ and $D$ are defined by
analogous expressions. Only the following combination of the two-point
functions is present:
$$ B_1(s) = B(s)-B(m_Z^2).  $$
$C_1(s)$ is the three-point function with one external massless and another
massive lines, {\it e.g.} $p_3^2=0$, $p_4^2=m_Z^2$, $(p_3+p_4)^2=s$.  The
expressions of the scalar loop functions in terms of Spence functions are
known \cite{dilog}.  $E(s,t)$ is the auxiliary function defined as:
$$E(s,t)=sC(s)+tC(t)+s_1C_1(s)+t_1C_1(t)-s t D(s,t).  $$
After routine manipulations helicity amplitudes expressed through the
$Z\g\g\g$ polarization tensor obtained in \cite{GM} can be reduced analytically
to our expressions (1)-(9).

\section{Numerial Results}

We first present different contributions to the polarized cross sections for
$\g Z$-pair production in monochromatic $\g\g$ collisions.
We consider the extreme cases of $\lambda{\g_1}\lambda{\g_2}=\pm 1$, i.e.
full circular polarization for the incoming photons. The cross section
is given by
$$
\frac{d\sigma_{\l_1\l_2}(s)}{dcos\theta} =\sum_{\l_3,\l_4}
\frac{\alpha^4 \cos^2\theta_W}{32\pi s\sin^2\theta_W}|A_{\l_1\l_2\l_3\l_4}|^2
(1-\frac{m_Z^2}{s}).
$$
We used the following numerical values of parameters $\alpha=1/128$,
$m_W=80.22$~GeV, $m_Z=91.173$~GeV. Contribution of fermionic loops is
calculated using the expressions for helicity amplitudes from \cite{fermi}. For
the masses of fermions we take the following values $m_u$ = $m_d$ = $m_s$ =
$m_c$ = $m_e$ = $m_\mu$ = $m_\tau=100$~MeV, $m_b=5$~GeV.  As is shown in
Figs.~1a,b $W$ loop contribution dominates at photon-photon collision energies
above 250~GeV. Cross section for equal initial photon helicities
$\l_{\g_1}\l_{\g_2}=1$ is about two times larger than that for opposite photon
helicities $\l_{\g_1}\l_{\g_2}=-1$. Also the cross section of transverse $Z_T$
boson production is at least two orders of magnitude larger than that for
longitudinal $Z_L$ boson production. Asymptotically cross section of $\g Z_T$
production in photon fusion is about 7 times larger than the cross section of
photon-photon scattering $\g\g\to\g\g$, calculated earlier \cite{aaaa}. The
reason is that $ZWW$ coupling is $\cos\theta_W/\sin\theta_W$ times larger  than
the photon $WW$ one, and a symmetry factor of $1/2$ is missing.

For calculating the realistic cross sections of $\g\g\to\g Z$ in PLC we assume
that 90\% electron (positron) beam longitudinal polarization
($\lambda_{e_{1,2}} =\pm 0.45$) and 100\% laser beam circular polarization
($\l_{\g_{1,2}}=\pm 1$) are achievable. We consider the following polarizations
of electron and laser beams: $2\l_{e_1}=2\lambda_{e_2}=0.9$ and
$\lambda_{\g_1}=\l_{\g_2}=-1$, which give the photon-photon energy spectrum
peaking just below the highest allowed photon-photon energy with mostly equal
photon helicities \cite{plc}.  Based on the $e^+e^-$ linear collider, PLC will
have almost the same energy and luminosity, i.e. c.m.  energy of $100-500$~GeV
and luminosity of order $10^{33}$cm$^{-2}$s$^{-1}$ are considered \cite{plc}.
In Fig.~2 are presented cross sections for $\g\g\to\g Z$ scattering for
longitudinal and transversal polarization of $Z$ boson as a function of the
$e^+e^-$ c.m. energy for polarized initial electron and laser beams. We
restrict $Z$ boson scattering angle, $|\cos\theta_Z|<\cos 30^\circ$. The
contributions of $W$ and fermionic loops as well as their sum are given
separately. Also a background from resolved photon contribution $q\bar q\to\g
Z$ is shown. We used the parametrization of photonic parton distributions from
\cite{grv}. As in Fig.~1, the cross section for transversely polarized $Z$
boson is two orders of magnitude larger than longitudinal one.  The resolved
photon contribution is larger than longitudinal $Z_L$ boson contribution, but
more than an order of magnitude smaller than dominating transverse $Z_T$ boson
contribution. At $e^+e^-$ collision energies above 250~GeV $W$-loop
contribution dominates. At $\sqrt{s_{e^+e^-}}=300$~GeV $\sigma_{\g\g\to\g
Z}=10$~fb, while at $\sqrt{s_{e^+e^-}}=500$~GeV $\sigma_{\g\g\to\g Z}=50$~fb,
which corresponds to 100-500 events for integrated luminosity of $\int{\cal
L}dt =10$~fb$^{-1}$. Unlike the case of hadron collider, PLC provides an
attractive environment not only for leptonic $Z\to ee,\ \mu\mu$, but also for
hadronic decay modes $Z\to q\bar q$. $\pi^0$ background events in jets $\g\g\to
q\bar q Z$ and continuum $\g\g\to q\bar q\g$ production must be rejected by
good geometric resolution and stringent isolation criteria combined with fair
jet-jet hadronic energy resolution to detect the $Z$ peak and accompanying
photon. The detection of both photon and $Z$ boson is also necessary to
suppress the background from the photon scattering off the residual electrons
left over from the original Compton backscattering $\gamma e\to Z e$.

Using the helicity amplitudes from Section~3 we calculated the $Z\to\g\g\g$
decay rate. The $W$ loop contribution of 0.026~eV coincides with the
calculation of \cite{GM}. For the total width we obtained 1.5~eV, which is
larger than calculated  in \cite{GM}. However this difference is due to another
value for $W$ mass used in \cite{GM}, because the fermionic contribution to
decay rate does depend  sensitively on the choice of
$\sin{\theta_W}^2=1-m_W^2/m_Z^2$.

\section{Conclusions}

We obtained the compact expressions for $W$ loop contributions to the helicity
amplitudes for $Z\g\g\g$ scattering in two different gauges ('t~Hooft-Feynman
and non-linear ones). As a result, we calculated cross sections of $\g\g\to\g
Z$ scattering at the Photon Linear Collider. Numerical values of the cross
sections indicate, that this process should be easily observable at the PLC.
For $e^+e^-$ c.m. energy of $300-500$~GeV or higher it will be possible to
separate  the $W$ boson loop contribution. Taking into account that $\g Z$
final state should be background free, we conclude that hundreds of $\g Z$
pair production events yearly should be observable at PLC. This fact will have
fundamental significance, because this process is a pure one-loop effect of the
Standard Model as a renormalizable nonabelian gauge theory. As such, the
observation of this reaction will provide a possibility to test triple and
quartic $\g WW$, $ZWW$, $\g\g WW$ and $\g ZWW$ vertices. If the anomalous
triple vertex, contributing mainly for transverse gauge bosons, {\it e.g.} so
called blind direction operator
$Tr\left(W_{\alpha\beta}W^{\beta\g}W_\g^\alpha\right)$ \cite{rujula}, exists
its effect can be probed in the reaction $\g\g\to\g Z$. For a composite $Z$
boson \cite{renard} one can also expect to observe a deviation from the
Standard Model predictions.

\vspace*{1cm}
We are grateful to S.S.~Gershtein and I.F.~Ginzburg for valuable discussions
and support. This work was supported, in part, by the Russian Foundation for
Fundamental Research.

\newpage
\section*{Figure captions}
\parindent=0pt
\parskip=\baselineskip

Fig.~1. Total cross section of $\g Z$ pair production in monochromatic
photon-photon collisions versus $\g\g$ c.m. energy for different helicities of
the incoming photons and final $Z$ boson. Total cross section (solid line) as
well as $W$ boson loop contribution (dashed line) and fermion loop contribution
(dotted line) are shown.

Fig.~2. Cross section of $\g Z$ pair production in $\g\g$ collisions versus
c.m. energy of the $e^+e^-$ collisions computed taking into account photon
spectrum of the backscattered laser beams. Dash-dotted line shows the resolved
photon contribution.  The other notations are the same as in Fig.~1.
\end{document}